\documentstyle[sprocl,epsf]{article}

\def\Journal#1#2#3#4{{#1} {\bf #2}, #3 (#4)}

\def\NPA{{\em Nucl. Phys.} A}
\def\NPB{{\em Nucl. Phys.} B}
\def\PLB{{\em Phys. Lett.}  B}
\def\PRL{\em Phys. Rev. Lett.}
\def\PRC{{\em Phys. Rev.} C}
\def\PRD{{\em Phys. Rev.} D}

\def\AP{{\em Ann. Phys.}}
\def\CJ{{\em Comp. J.}}

\def\be{\begin{equation}}
\def\ee{\end{equation}}
\def\bea{\begin{eqnarray}}
\def\eea{\end{eqnarray}}

\newcommand{\ben}{\begin{eqnarray*}}
\newcommand{\een}{\end{eqnarray*}}
\newcommand{\drms}{DR[$\overline{\mbox{MS}}$]}

\begin{document}

\title{COMPARISON OF REGULARIZATION METHODS FOR NUCLEON-NUCLEON
EFFECTIVE FIELD THEORY}

\author{JAMES~V.~STEELE~\footnote{Email: jsteele@mps.ohio-state.edu}}

\address{Department of Physics, Ohio State University\\ 
	 Columbus, OH 43210-1168}


\maketitle\abstracts{
The characteristics of a meaningful effective field theory (EFT)
analysis are discussed and compared with traditional approaches
to $NN$ scattering.  A key feature of an EFT treatment is a
systematic expansion in powers of momentum, which is demonstrated using 
an error analysis introduced by Lepage.  A clear graphical
determination of the radius of convergence for the momentum expansion
is also obtained.  I use these techniques to compare cutoff
regularization, two forms of dimensional regularization, and the  
dibaryon approach, using a simple model for illustration.   The
naturalness of the parameters and predictions for 
other observables are also shown.
}

\section{Introduction}

An effective field theory (EFT) description of nucleon-nucleon ($NN$)
scattering is an important step on the road to a consistent
description of nuclear matter.  It also provides the techniques to
predict other observables such as bound state expectation values.
However, the interaction of two heavy nucleons requires a 
nonperturbative treatment~\cite{Weinberg,BiraThesis} which has lead to
disagreements about the nature and limitations of an EFT
expansion in this 
case.~\cite{Maryland,Lepage,KSW1,LukeManohar,Kaplan,Richardson,Gege}
Regularization is needed to handle divergences that arise, but the
results are said to depend on the regularization scheme used and the
size of the scattering length involved. 
More generally, it is claimed that the behavior and predictive power
expected from a true effective field theory is not exhibited by every
regularization method when applied nonperturbatively.~\cite{Maryland,Lepage}

Therefore, R.~J.~Furnstahl and I
set out to determine the
important features of a nonperturbative EFT\footnote{All the work
presented here has been done in collaboration with
R.~J.~Furnstahl.~\cite{OurPaper}}.
This can be done most
clearly by making a side-by-side comparison of the regularization
schemes listed in Table~\ref{tab1} using the error analysis advocated by
Lepage.~\cite{Lepage} This comparison illustrates which
schemes behave like a true EFT,~\cite{OurPaper} as will be seen
below.   In addition, the differences seen from a comparison with
successful phenomenological models for $NN$ scattering such as the
Reid potential~\cite{Reid} help determine the importance of
a reformulation in terms of an EFT.

\begin{table}[t] \caption{\label{tab1} Regularization schemes
and their abbreviations used throughout this paper.}
\begin{tabular}[tbh]{|cp{3.83in}|}
\hline 
name & \hfill regularization scheme \hfill\null \\ \hline
CR[G] & Cutoff regularization with gaussian weighting in the 
potential~\cite{Lepage}
\\
\drms{} & Dimensional regularization with modified minimal 
subtraction~\cite{KSW1}
\\
DR[PDS] & Dimensional regularization with power divergence
subtraction~\cite{KSW2} 
\\
dibaryon & \drms{} but with an additional 
       degree of freedom associated with
         a low-energy bound or nearly bound state~\cite{Kaplan,vanKolck}.
\\ \hline
\end{tabular}
\end{table}

\section{Comparison Techniques}

I first describe how to analyze and compare the different
regularization schemes in a way that will allow for a determination of which
schemes have the 
systematics and predictability desired of an effective field theory.
Concentrating on $S$-wave scattering, I will focus
only on the short-distance EFT.   This general $S$-wave
potential can be written as 
\be
V({\bf p},{\bf p'}) = \frac{4\pi}{\Lambda_s^2} \left( c - d\;
\frac{{\bf p}^2+{\bf p'}^2}{2\Lambda_s^2} + e\; 
   \frac{({\bf p}^2+{\bf p'}^2)^2}{4\Lambda_s^4}
+ \ldots \right) \ ,
\label{eq:pot}
\ee
with a scale $\Lambda_s$ introduced to make the coefficients $c$, $d$,
$e$, $\ldots$, dimensionless.  There is also a scale $\Lambda$
implicit in this effective potential that signifies the point above
which new degrees of freedom not accounted for in the effective theory
begin to contribute to the physics.~\cite{Lepage}
The factor of $1/4\pi$ picked up by
each additional 
term in the Born series requires a $4\pi$ to be factored out in order
to render these coefficients natural,~\cite{Weinberg} {\it i.e.\/} of
order unity, as discussed further below.  

The coefficients are determined order-by-order from matching to the
available data. To fit the first $n$ constants in the potential
requires at least $n$ points of data.
The potential is not a measurable
quantity, so instead  
a scattering observable such as the phase shift $\delta(p)$ is used to
determine these constants.  

Inserting this potential into the
Lippmann-Schwinger equation and specifying a regularization
scheme, the resulting amplitude determines the
phase shift by one of the following two equivalent relations
($T=-{\cal A}$) 
\be
T(p)=-\frac{4\pi}{Mp} e^{i\delta(p)} \sin\delta(p) \ ,
\qquad\qquad
-\frac{4\pi}{M} \frac1{T(p)} = p\cot\delta(p) - ip \ .
\label{eq:pcotdel}
\ee
For example, dimensional regularization with modified minimal
subtraction (\drms{}) gives
\be
T_{\rm DR[\overline{MS}]} = \frac{V(p,p)}{\displaystyle
1+\frac{iMp}{4\pi} V(p,p)} \ .
\ee
The imaginary part of the denominator is linked to the numerator by 
unitarity.  In general, however, the real part of the denominator can
be more complicated than in the \drms{} case and consequently dictates
the behavior of the scheme near a bound state.  
This can be seen by calculating the same quantity for dimensional
regularization with power divergence subtraction (DR[PDS]) 
\be
T_{\rm DR[PDS]}  = \frac{V(p,p)}{\displaystyle
1+\frac{M}{4\pi}(\mu + ip) V(p,p)} 
\label{eq:Tpds}
\ee
which reduces to the \drms{} result for $\mu=0$.  However, the
appearance of this additional mass scale is important for a good radius
of convergence as will be seen in the next section.  

Cutoff regularization is another method for dealing
with divergences in which momenta greater than the scale of new
physics $\Lambda$ are explicitly suppressed.   This can be
implemented by adding a gaussian weight $\exp(-({\bf p}^2+{\bf
p'}^2)/2\Lambda_c^2)$ to the potential in 
Eq.~(\ref{eq:pot}) and will be referred to as CR[G].  Although a
physically intuitive method, numerical
techniques are required 
to solve this type of regularization.  However, one can see the
effect of the cutoff by taking a geometric sum of the
first two Born terms of the 
CR[G] method.  This just ends up giving Eq.~(\ref{eq:Tpds}) with
$\mu=2\Lambda_c/\pi$. 
It will be shown below that \drms{} has very different results from both
DR[PDS] and CR[G], which can be traced to this difference in the real
parts of the $T$-matrix denominators.

\begin{figure}
\begin{center}
\leavevmode
\hbox{
\hspace{-1.25cm}
\epsfxsize=2.75in
\epsffile{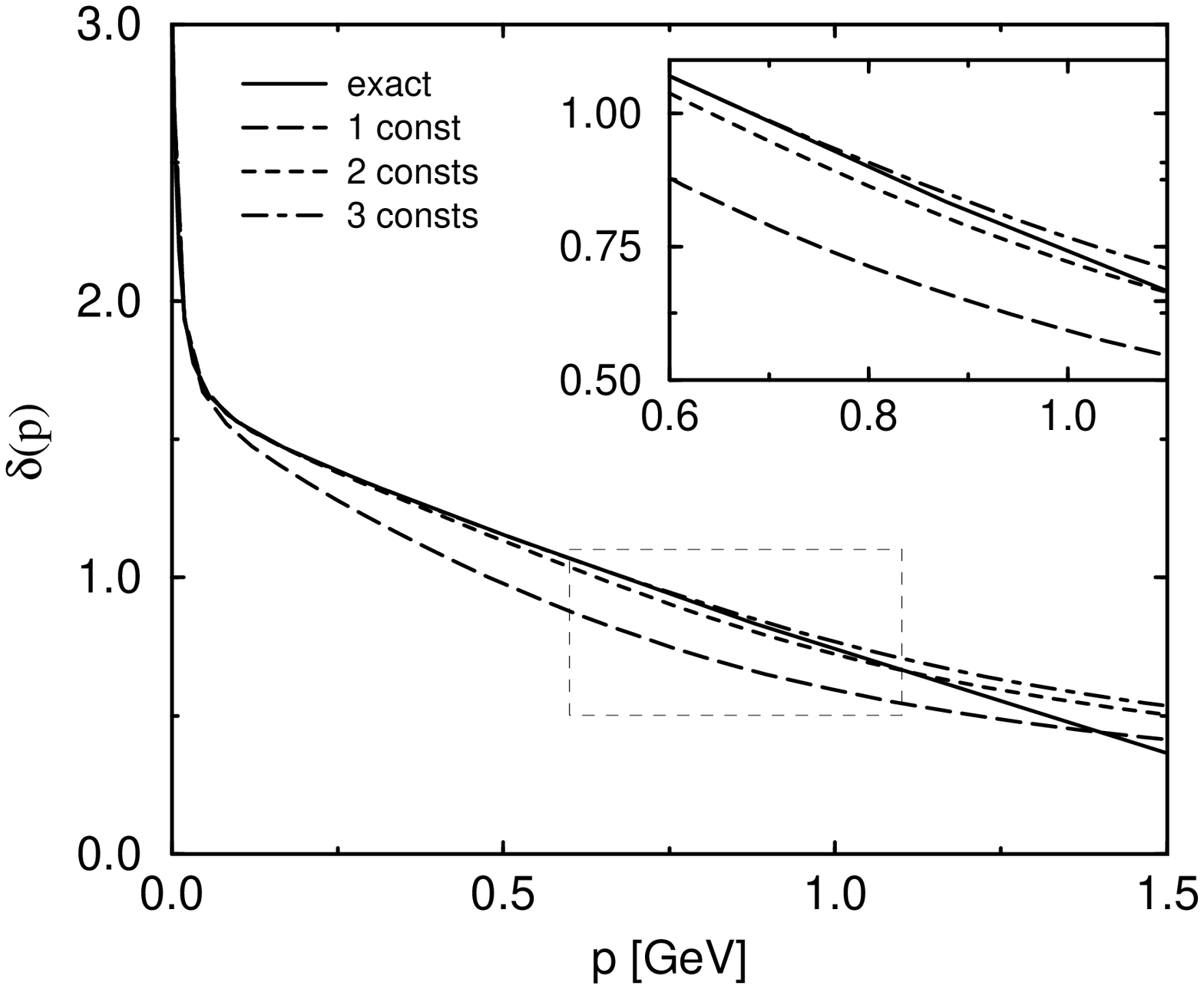}
\hspace{-.5cm}
\epsfxsize=2.75in
\epsffile{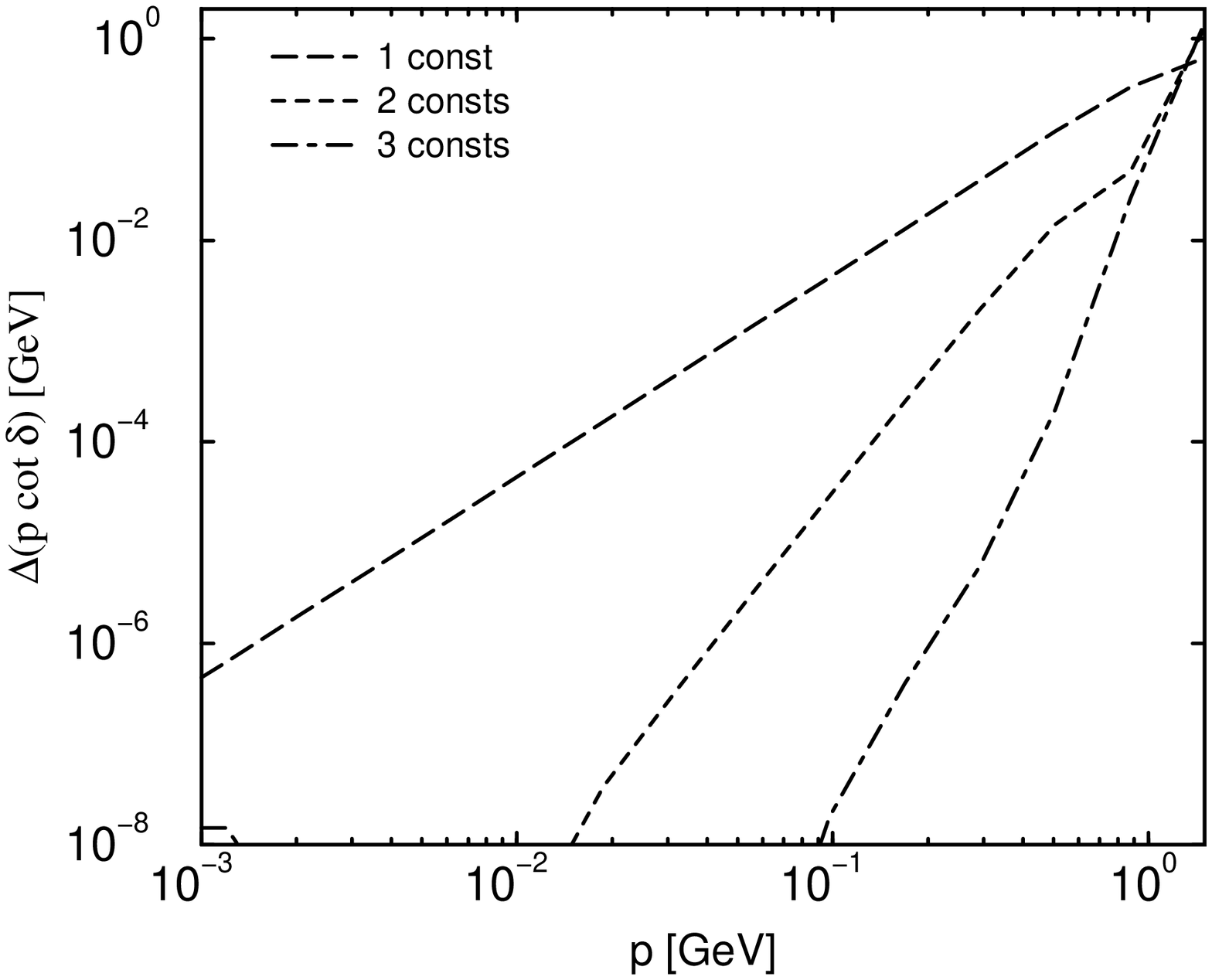}
}
\end{center}
\caption{\label{phaseshift}The phase shift $\delta(p)$ (left) and the
error in $p\cot\delta(p)$ (right), each plotted as a function of
$p$ for the delta-shell potential with a weakly bound state,
as discussed in Section~3.  The solid line is the exact result
and the dashed lines show the CR[G] fit for one, two, and three
constants.}
\end{figure}

In the left plot of Fig.~\ref{phaseshift}, I show the results from
using CR[G] to 
fit one, two, and three constants in the potential,
as compared to the exact $S$-wave phase-shift for the delta-shell
potential,~\cite{Gottfried} which models the underlying physics and is
discussed in detail in the next section.
A first glance at the plot shows the approximation to the phase shift
improves as more constants in the potential are fit.
However, a second look shows that it is very difficult to gather
any {\it quantitative\/} information from the plot.  At what point the curve
deviates enough to be considered inaccurate is
not clear, and the radius of convergence
of the EFT expansion is completely obscure.  Furthermore, 
a simple calculation shows that every term in the momentum
expansion of the phase shift contains the scattering length.~\cite{KSW1} 
This means that the error in the phase shift could be
numerically sensitive to a large scattering length and contaminate the power
counting.

This issue is avoided with no loss of generality
by plotting the error of $p\cot\delta(p)$ instead:
\be
\Delta [ p\cot\delta(p) ] \equiv | p\cot\delta_{\rm eff}(p) -
p\cot\delta_{\rm true}(p) | \ .  
\label{eq:seven}
\ee
If the effective field theory follows proper power counting, then this
error should improve by two powers of momentum as each additional
coefficient in the potential is fixed in Eq.~(\ref{eq:pot}).
Since a large scattering length is synonymous
with a near bound state (or pole) in the amplitude, Eq.~(\ref{eq:pcotdel})
shows that this pole is cleanly mapped only to the first constant in a
momentum expansion of $p\cot\delta(p)$.  
It is known from
conventional scattering theory that this combination has a well defined
expansion in $p^2$ for short-range potentials known as the ``effective
range expansion'' 
\be
p\cot\delta(p) = -\frac1{a_s} + \frac12 r_e p^2 + v_2 p^4 + \ldots \ ,
 \label{eq:ere}
\ee
which defines the scattering length $a_s$ and effective
range $r_e$.  When long-range potentials are included, 
Eq.~(\ref{eq:ere}) is only valid at  low momentum
or is even inapplicable. 
In this case, one must define a modified effective range
expansion.~\cite{berger}  
Since the effective theory contains the same long-distance physics as
the true underlying theory, Eq.~(\ref{eq:seven}) can be modified to
have a clean momentum expansion.
For short range potentials as considered here, Eq.~(\ref{eq:seven}) is
sufficient.

Plotting the error in $p\cot\delta(p)$ as a function of $p$ on
a log-log plot,  a straight line is expected with slope given by the dominant
(lowest) power of $p/\Lambda$ in the error.~\cite{Lepage}
As more constants are included, the slope in this error should increase,
signifying the removal of higher powers of $p/\Lambda$.  The second
plot of Fig.~\ref{phaseshift} clearly demonstrates 
the order-by-order improvement
in the amplitude as more constants are added to the effective potential.
With one constant the slope of the error is two, {\it i.e.}, 
${\cal O}(p^2/\Lambda^2)$, and increases by two
with each additional constant. 

\section{Illustration with Delta-Shell Potential}

In this section, I illustrate the points made above by using EFT
techniques with the different regularization methods to systematically
describe the ``unknown'' short-range physics of a specific example.
I could use $NN$ scattering data, but it is more
convenient to use an exactly solvable potential to serve as data in
order to have a clean understanding of what features are important.  
The delta-shell
potential has been used in the past to simulate the large scattering
length found in $NN$ scattering.~\cite{Gottfried}  
Kaplan used
this potential to illustrate the benefits of the dibaryon
approach.~\cite{Kaplan}  He found upon considering $NN$ scattering
that the 
inclusion of long-distance pion physics did not change the
conclusions.  This agrees with the
experience of other authors~\cite{Maryland,KSW1} that the addition
of pions as long-range interactions does not affect the properties
of the short-range expansion.
The delta-shell potential
is therefore a sufficient model for the purposes here.

The potential can be written in terms of the nucleon mass $M$, the
coupling $g$, and the range of the potential $r_0$.  This short-range
potential represents the new physics of the underlying model theory,
and so $r_0=1/\Lambda$ is taken below,
\be
V_{\rm true}(r) = -g \frac{\Lambda}{M}\,
\delta\left(r-\frac1\Lambda\right) .  \label{thispotential}
\ee
It has exactly one bound state for $g\ge1$ and no bound states
for $g<1$.  Scattering with $p >
\Lambda$ probes the details of the potential, so one would 
expect $\Lambda$ to be the radius of convergence of a well-tuned EFT.
The scattering length becomes very large for $g$ near 1,
whereas the effective range (and the rest of the terms in the
effective range expansion) are of natural size for all $g$:
\be
a_s = \frac{g}{g-1}\; \frac1{\Lambda}\ ,
\qquad\qquad
r_e =  \frac{2(g+1)}{3g} \; \frac1{\Lambda}\  .
\label{scat}
\ee
The scattering length in 
the ${}^1S_0$ channel of $NN$-scattering can be
modeled by choosing
$(g,\Lambda)=(0.99,m_\rho)$.  
This potential Eq.~(\ref{thispotential}) with different $g$'s will be the
``laboratory''  from which  the different regularization
schemes are compared.  I will also briefly discuss results for actual measured
data and the inclusion of pions below.

First, some intuition will be gained by graphically reproducing Kaplan's
result that \drms{} has a small radius of convergence if the
scattering length is large. From Eq.~(\ref{scat}), it can be seen that
choosing $g=0.99$ gives a scattering length one 
hundred times
larger than choosing $g=-10$.  At the same time,  the
effect from the presence of a bound state is investigated by taking $g=1.01$. 
The effective potential is given by
Eq.~(\ref{eq:pot}) with the mass scale $\Lambda_s$ associated
with the inverse delta-shell radius and the prescription of using \drms{} on
all divergent integrals.

\begin{figure}
\begin{center}
\leavevmode
\hbox{
\hspace{-1.25cm}
\epsfxsize=2.75in
\epsffile{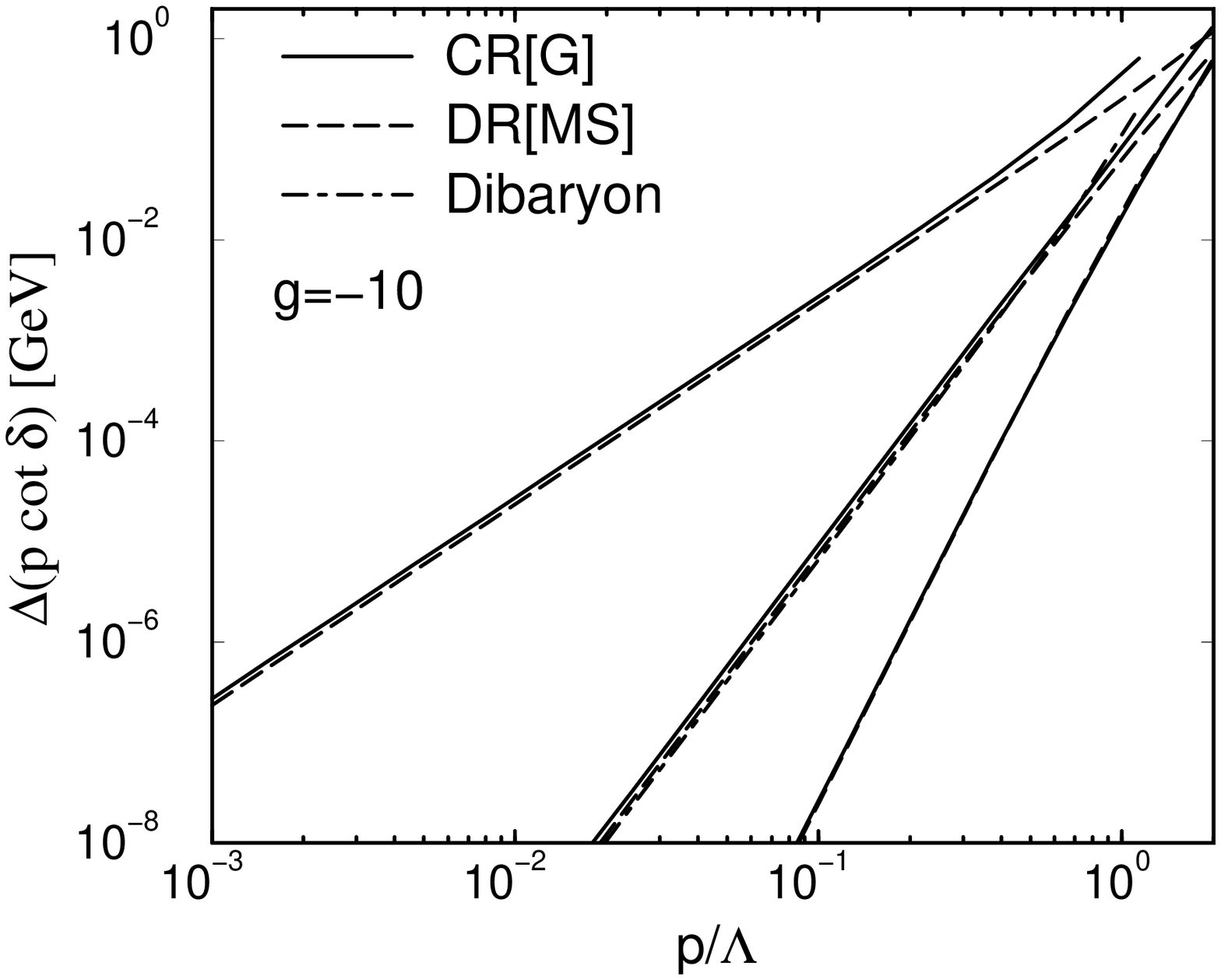}
\hspace{-.5cm}
\epsfxsize=2.75in
\epsffile{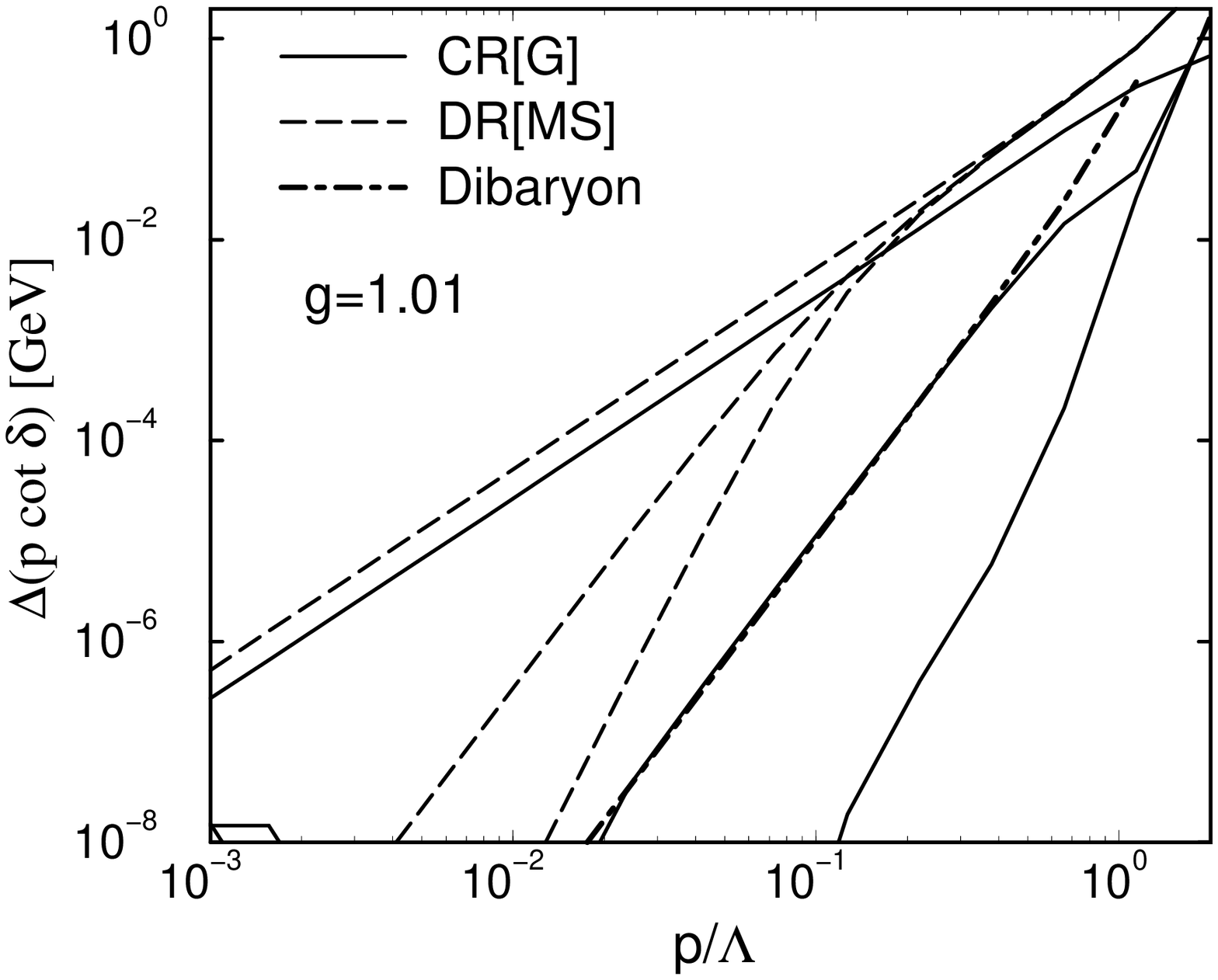}
}
\end{center}
\caption{\label{err1}The error in $p\cot\delta(p)$ plotted as a function of
$p/\Lambda$ for a small scattering length without a bound state $g=-10$
and for a large scattering length with a bound state $g=1.01$.}
\end{figure}

The momentum expansion for \drms{} can be shown
analytically~\cite{KSW1,Kaplan} to be $p^2 a_s r_e/2$.  This implies
using Eq.~(\ref{scat}) that the radius of convergence for $g=1.01$ and
$0.99$
should be roughly $1/10$ that of the $g=-10$ case.  Fixing the
constants in the potential Eq.~(\ref{eq:pot}) by matching
$p\cot\delta(p)$  produces the
results in Fig.~\ref{err1}.  The dashed 
lines show the \drms{} results for one, two, and three constants
respectively.  Indeed all three lines converge to $p/\Lambda\sim1$ for
$g=-10$ and $p/\Lambda\sim0.1$ for $g=1.01$.  The results for $g=0.99$
fall on top of the $g=1.01$ results and are therefore not shown.  This
implies the presence of a bound state does not matter, but the size of
the scattering length does.  
It would be difficult to
draw these conclusions had  only the phase shift itself  been plotted
(the left plot of Fig.~\ref{phaseshift}).  Of course, 
these results can also be shown
analytically for this simple model, 
but the analysis applies much more generally, when part or all of the
calculation is done numerically.

The constants $c$, $d$, and $e$ are given in Tables~\ref{tab2} and
\ref{tab3}.  Their
values are required to at least 8 digits to produce the
accuracy of Fig.~\ref{err1}. The 
correlation between the naturalness of the constants and the radius of
convergence is apparent.  
The constants for the $g=1.01$ case are extremely large,
reflecting the breakdown of the effective field theory for \drms{} much
below the expected scale $\Lambda$.

\begin{table}[t] 
  \caption{\label{tab2}Effective potential for $g=-10$ (small scattering
    length) to three different orders in the error ${\cal
    O}(p^n/\Lambda^n)$ for different regularization schemes.  $\Lambda
    a=1$ for CR[G] and $\mu=\Lambda$ for DR[PDS].}
\begin{tabular}{|c|ccc|ccc|ccc|}
\hline
error & \multicolumn{3}{c|}{\drms{}} & \multicolumn{3}{c|}{DR[PDS]} &
        \multicolumn{3}{c|}{CR[G]} \\  
order   & $c$ & $d$ & $e$ & $c$ & $d$ & $e$ & $c$ & $d$ & $e$ \\  \hline
2 & 0.758 & --- & --- &
                            8.33  & --- & --- &
                            1.49  & --- & ---  \\
4 & 0.758 & $-0.206$ & --- &
                            8.33  & $-25.0$  & --- &
                            2.67  & $-1.04$ & ---  \\
6 & 0.758 & $-0.206$ & 0.0672\ &
                            8.33  & $-25.0$  & 38.2\ &
                            2.67  & $-1.58$ & 3.75\  \\
\hline
\end{tabular}
  \caption{\label{tab3}Effective potential for $g=1.01$ (large scattering
    length) to three different orders in the error ${\cal
    O}(p^n/\Lambda^n)$ for different regularization schemes.  
	$\Lambda a=1$ for CR[G] and $\mu=\Lambda$ for DR[PDS].}
\begin{tabular}{|c|c@{\ \ }c@{\ \ }c@{\ }|%
c@{\ \ }c@{\ \ }c@{\ }|c@{\ \ }c@{\ \ }c@{\ }|}
\hline
error & \multicolumn{3}{c|}{\drms{}} & \multicolumn{3}{c|}{DR[PDS]} &
        \multicolumn{3}{c|}{CR[G]} \\  
order   & $c$ & $d$ & $e$ & $c$ & $d$ & $e$ & $c$ & $d$ & $e$ \\  \hline
2 & 84.2 & --- & --- &
                            $-0.842$  & --- & --- &
                            $-1.42$  & --- & ---  \\
4 & 84.2 & $-5640$ & --- &
                            $-0.842$  & $-0.564$ & --- &
                            $-0.946$   & $0.842$ & ---  \\
6 & 84.2 & $-5640$ & $378000$ &
                            $-0.842$  & $-0.564$ & $-0.152$ &
                            $-0.937$   & $0.618$ & $-0.181$  \\
\hline
\end{tabular}
\end{table}

One way to fix this behavior is to introduce the
dibaryon.~\cite{LukeManohar,Kaplan} This takes the large scattering
length into 
account by explicitly introducing a low-energy $s$-channel degree of
freedom into the effective lagrangian.  For $g>0$ or $g<-1$, 
the potential can be written as
\be
V_{\rm dib.}({\bf p},{\bf p'}) = C - \frac{y^2}{E+\Delta};
\mbox{\ \ }
C = \frac{2\pi}{M\Lambda},
\mbox{\ \ }
y^2 = \frac{3\pi\Lambda}{M^2} \frac{1+g}{g},
\mbox{\ \ }
\Delta = \frac{3\Lambda^2}{2M} \frac{1-g}{g},
\label{dibar}
\ee 
with $E=p^2/M$ always kept on-shell.  Since there seem to be three
constants fit in Eq.~(\ref{dibar}), one might think the dibaryon will
have an error of ${\cal O}(p^6/\Lambda^6)$.  However, the dibaryon
amplitude is only matched to second order~\cite{Kaplan} in the momentum when
deriving the relations in 
Eq.~(\ref{dibar}) and indeed shows an error of ${\cal
O}(p^4/\Lambda^4)$ for the two values of $g$ in Fig.~\ref{err1}
(plotted as the dot-dashed line).  The  
slope and magnitude of the error do not depend on the scattering
length, as expected.

I next repeat the calculations using the cutoff regularization method
CR[G] with $\Lambda_c=\Lambda$.  There are various ways to numerically solve
the Schr\"o\-dinger equation with a cutoff, but
the following procedure is particularly efficient
and numerically robust enough to attain the accuracy required here.  
First, the variable phase method is used to
solve for the phase shift.
This is a differential equation,
\be
\delta'(r) = - \frac{M}{p}\; V(r)\; \sin^2\!\left( pr + \delta(r)
\right)
\qquad\mbox{with\ } \delta(0)=0 \ ,
\ee
which expresses the change in the phase shift as
the potential is built up from zero at $r=0$ to its full value at
$r=\infty$.  
The boundary condition ensures that the full phase shift given by
$\delta(\infty)$ is zero in 
the absence of a potential and defines the otherwise ambiguous
multiple of $\pi$ in the phase shift. The routine ODE from
package ODE~\cite{ODE} was used to solve the differential equation.  

Then I use a general method for taking the theoretical error of the
EFT into account when matching the constants in the effective
potential. After evaluating the combination
$p\cot\delta_{\rm eff}(p)$ in the effective theory, it is subtracted
from the true result (either data or an exact solution to a model
problem) and the difference,
\be
\Delta p\cot\delta(p) = \alpha + \beta \frac{p^2}{\Lambda^2} + \gamma
\frac{p^4}{\Lambda^4} + \ldots   \ ,
\label{diffER}
\ee
is fit to a polynomial in
$p^2/\Lambda^2$ to as high an order as possible.  
To obtain the accuracy shown in the plots, a weighted
polynomial fit of $\Delta p\cot\delta(p)$ up to $p/\Lambda=0.1$ was
required.

Using a spread of momentum near zero, the polynomial fit should be
weighted with both the expected theoretical error in momentum and any
additional experimental noise.  The resulting coefficients $\alpha,
\beta, \gamma, \ldots$ are then minimized with respect to variations
in the effective potential constants $c,d,e,\ldots$ using an
optimization code.  In practice, this method is more robust and
numerically stable than matching the values of $p\cot\delta(p)$ at
discrete points to fix the constants. This allows the analysis of
Lepage~\cite{Lepage} to be extended beyond second order.  Also note
that such a procedure is needed when matching to data even when the
EFT observables can be calculated analytically.

The number of coefficients that can be minimized is given by
the number of constants retained in the effective potential.  
I used DPOLFT from package SLATEC~\cite{ODE} to find the polynomial fit
and MINF,~\cite{MINF} which is based on the Powell method, to carry out the
minimization.  Normal accuracies in minimization using double
precision numbers with this method are $10^{-12}$ or better. 

The solid lines in Fig.~\ref{err1} show that CR[G] does work
regardless of the scattering length.  In fact, 
with $\Lambda_c \sim \Lambda$, the result is just as
good as the dibaryon for the same number of constants.  The values for
the constants are given in Tables~\ref{tab2} and \ref{tab3}, showing
they are all 
natural for both $g$'s (although the third constant is somewhat 
small for $g=1.01$).

Note that as more constants are fixed, the lower-order constants are
modified.  
This occurs because even after truncating the potential for the cutoff
regularization to a given order, it still contains all orders in
$p^2$ from the gaussian factor.
The nonperturbative solution of the Lippmann-Schwinger equation
therefore can generate terms of {\em any} order in $p^2/\Lambda^2$.
The amplitude itself is matched to the true result order-by-order in
the momentum so the power counting of the potential is destroyed.
This consequence of the cutoff regularization is not necessarily
relevant since
the potential is not an observable. 
It is interesting to note, however, that these
modifications are relatively small. This is also true when adding the long
distance physics of the pion in fitting to actual $NN$
data as shown in the next section.

In summary, Fig.~\ref{err1} shows that all regularization
schemes considered so far  produce useful effective field theories for
a small scattering length, but \drms{} fails for large scattering
length.  A failure of the power counting in powers of $p/\Lambda$ is
reflected in unnatural constants in the potential.

\begin{figure}
\begin{center}
\leavevmode
\hbox{
\hspace{-1.25cm}
\epsfxsize=2.75in
\epsffile{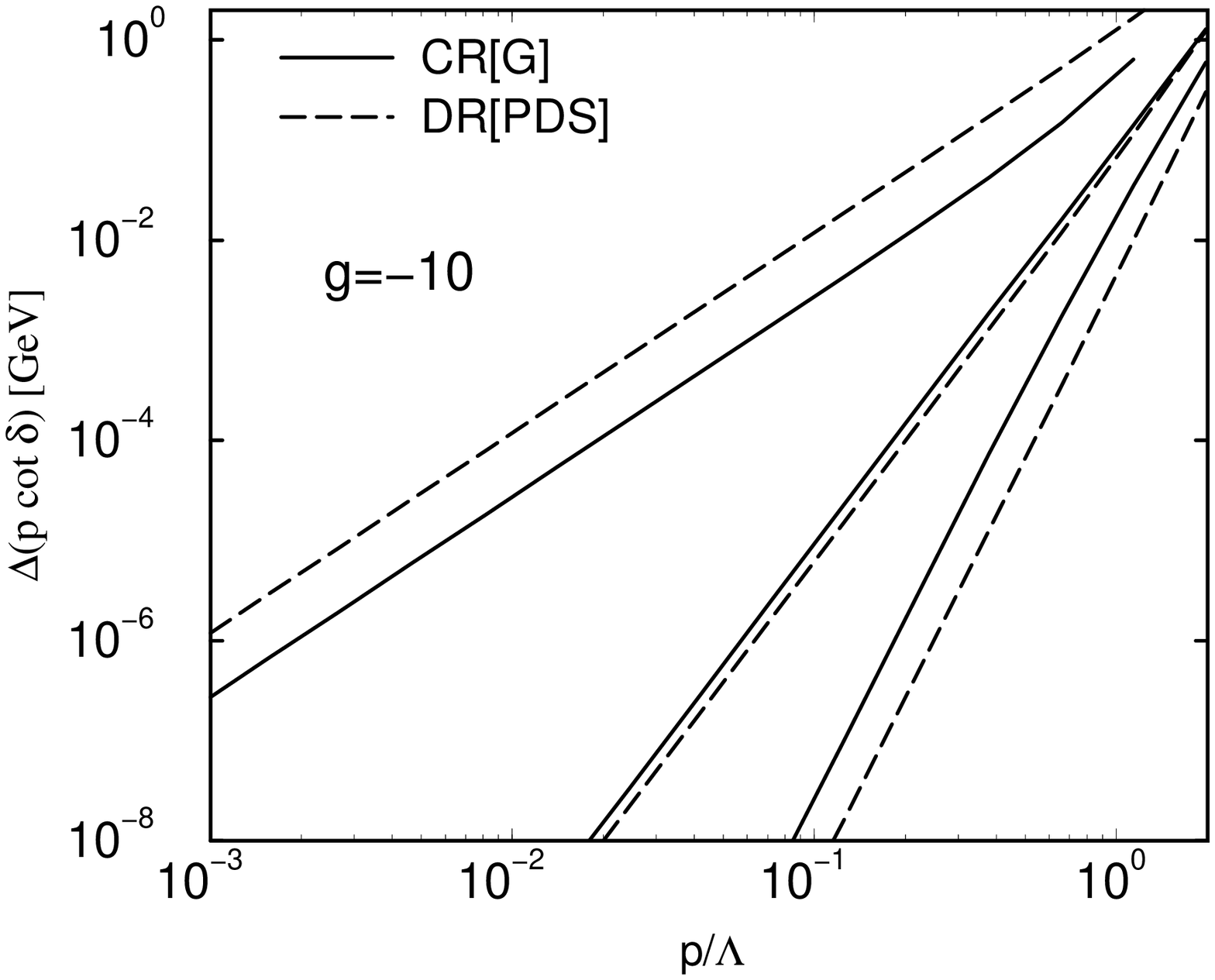}
\hspace{-.5cm}
\epsfxsize=2.75in
\epsffile{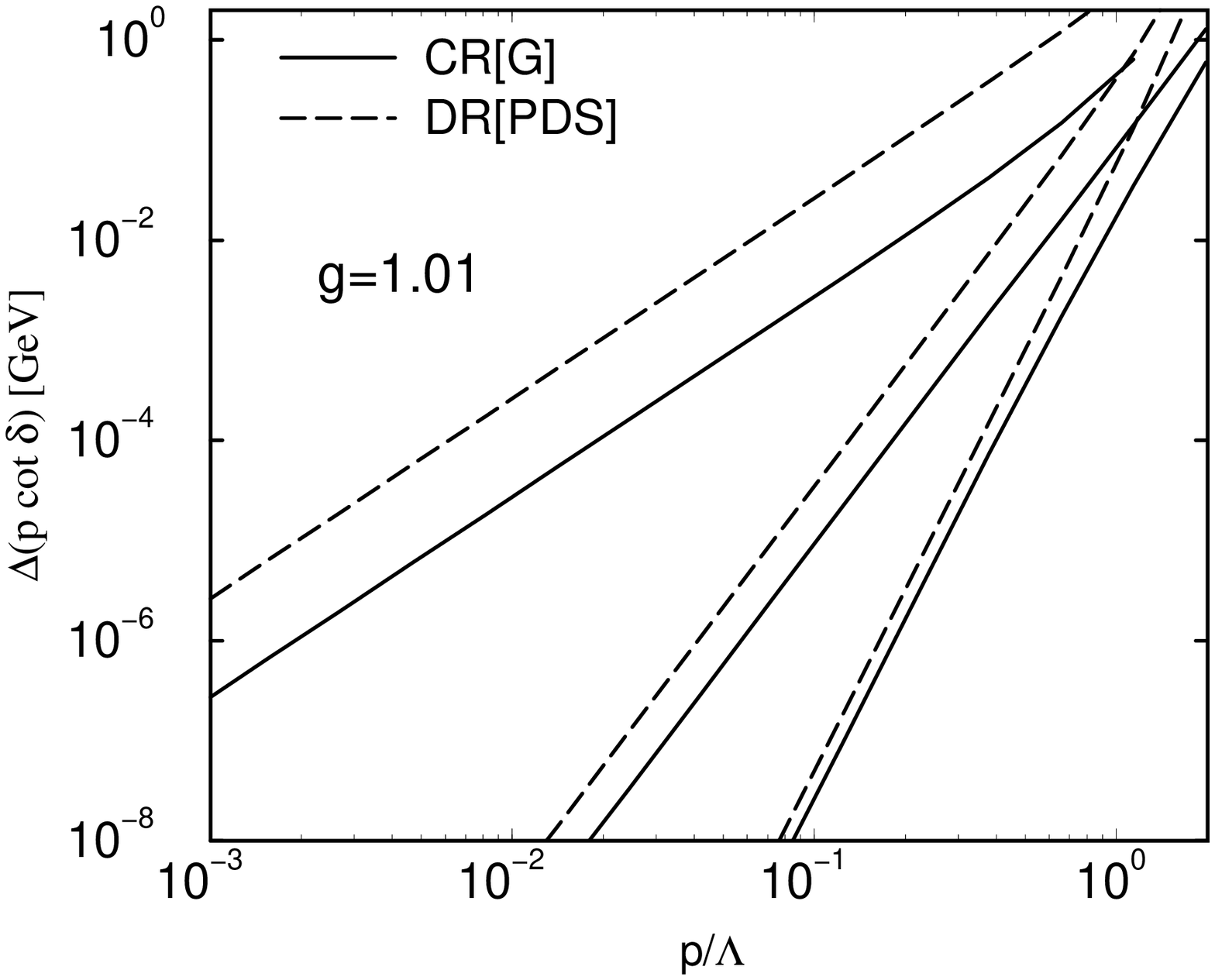}
}
\end{center}
\caption{\label{err2}The error in $p\cot\delta(p)$ plotted as a function of
$p/\Lambda$ for a small scattering length without a bound state $g=-10$
and for a large scattering length with a bound state $g=1.01$.}
\end{figure}

I now turn to the most recently proposed regularization
scheme,~\cite{KSW2} DR[PDS].
An additional prescription compared to the \drms{} case is an
expansion of observables to the same order in $p^2$ as the potential
Eq.~(\ref{eq:pot}). 
If this is not done, the results are $\mu$ dependent, with
$\mu=0$ reproducing \drms{} and $\mu$
larger than the nucleon mass approaching the 
CR[G] result in Fig.~\ref{err1}. 

When only dealing with a short-range potential as discussed so far,
the DR[PDS] 
prescription reproduces the effective range expansion Eq.~(\ref{eq:ere})
by construction. 
The DR[PDS] results in Fig.~\ref{err2}  
are therefore $\mu$ independent, but the constants are still $\mu$
dependent. I take $\mu=\Lambda$ to compare with CR[G]. 
 This produces natural constants for
$g=1.01$ but somewhat unnatural ones for $g=-10$ as seen in
Table~\ref{tab2}.  This is an accidental consequence of the momentum
expansion being in powers of $p^2 r_e/2(\mu-1/a_s)$, so that for $g=-10$
and $\mu=\Lambda$, Eq.~(\ref{scat}) shows the
denominator is nearly zero.  
This implies that the scale $\mu$ is not 
functionally equivalent to the cutoff
$1/a$ in CR[G] since it does not always signal the onset of new
physics at the scale $\Lambda$.

For both large and small scattering length, DR[PDS] does quite
well, with a radius of convergence $p/\Lambda\sim 1$.  The CR[G]
result is better for one constant since the cutoff generates
an effective range $r_e$ close to the true result.  Overall,
DR[PDS] is a convenient method to produce reasonable
analytical results, and depending on the problem at hand either CR[G]
or DR[PDS] may be suitable.  
One should note, however, that only DR[PDS] provides a strict diagram
by diagram power counting.~\cite{KSW2}

Finally, I mention the Reid potential~\cite{Reid} 
which consists of a sum of
Yukawa interactions, with fixed masses chosen as integer multiples of the
pion mass, and coefficients that can be varied to fit the data.
So for $S$-waves,
\bea
V_{\rm Reid}({\bf p},{\bf p'}) &=& \frac{c_1}{{\bf q}^2+m_\pi^2} 
+\frac{c_2}{{\bf q}^2+(3m_\pi)^2} 
+\frac{c_3}{{\bf q}^2+(5m_\pi)^2} + \ldots 
\label{reidpot} \\
&=& {\rm OPE} + \frac{c_2 (5m_\pi)^2 + c_3 (3m_\pi)^2 + (c_2+c_3) {\bf
q}^2}{[{\bf q}^2+(3m_\pi)^2][{\bf q}^2+(5m_\pi)^2]} \ .
\nonumber
\eea
Combining the Yukawa terms in the second equation generates a
numerator that looks like an EFT and a denominator 
that suppresses large momentum, similar to the cutoff regularization.
However, since each term has a different mass,
a clear separation scale $\Lambda$ is not identified.  
There is also a one-pion exchange (OPE) term which is not important for the
short-range physics discussed here.

The original Reid analysis used a
global fit and only one adjustable parameter in each channel,
for a result with
approximately the same error (roughly a few percent of the data) at
all momenta.
However,  the EFT fitting procedure can be applied instead.  
Using Yukawa
masses comparable to $\Lambda$ should give similar results
to CR[G].
Indeed, if the OPE contribution is ignored and a low-momentum fit is
done to the constant 
$c_2$, the error plot is
similar to the CR[G] result with one constant (Fig.~\ref{err1}).
Adding a second short-range Yukawa does as well as CR[G] with two
constants, since the Yukawas play off each other to allow the next order
error in ${\bf q}^2$ to be removed.
This interplay becomes increasingly complex at higher orders.
Furthermore,
the additional mass scales  obscure (or smear out) 
the role of
$\Lambda$ as a  scale that separates the known from the
unknown physics in effective field theories.
Traditional $NN$ potentials such as Reid 
are well suited for global fits, but
systematic predictions with controlled error estimates are more
properly analyzed using
an effective field theory.

\section{Other Observables}

I now turn to an investigation of the binding energy.  If the EFT is
truly reproducing the S-matrix of the underlying theory order-by-order
in a momentum expansion, it should reproduce the binding energies and
other observables to the same order of accuracy as the phase shifts.
I therefore use the binding energy prediction as a consistency check
for the candidate effective field theories found so far.

The potentials have already been fit to a given order by the scattering
phase shifts above, and I use these potentials 
without adjustment to solve for the
binding energy. This can be done analytically for the DR schemes by
finding the poles in the scattering amplitude.
The exact binding energy $E_{\rm bind}$
for the delta-shell potential is given by solving the equation
\be
\frac1{g} = \frac{1-e^{-2\eta}}{2\eta},
\qquad\qquad
\eta=\frac{\sqrt{M E_{\rm bind}}}{\Lambda} .
\ee
There is only one bound state to predict in the delta-shell
potential, and if it is shallow enough, even the effective range
expansion with the values of $a_s$ and $r_e$ can determine its value.
A better test is to increase $g$ until $E_{\rm bind}$ is large and
on the order of $\Lambda$, and use the EFT to determine the accuracy of the
binding energy prediction as a function of this variation.  
If a true radius of convergence is present, the effective field theory
should break down for $E_{\rm bind}/\Lambda\sim 1$.  The binding energy
is $6.48\times 10^{-2}$ MeV for $g=1.01$ but quickly increases
to $812$ MeV for $g=2.5$. 

\begin{figure}
\begin{center}
\leavevmode
\hbox{
\hspace{-1.25cm}
\epsfxsize=2.75in
\epsffile{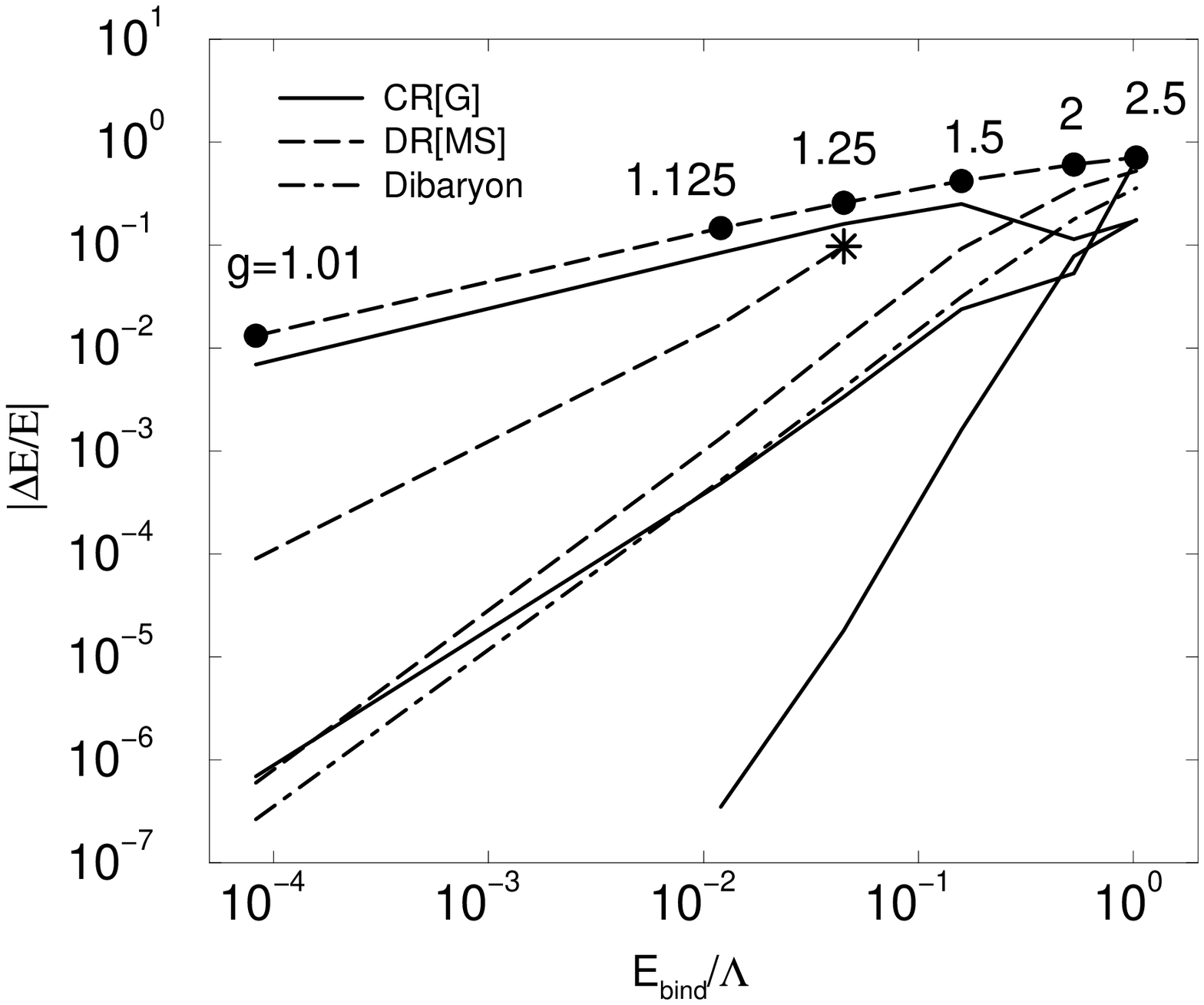}
\hspace{-.5cm}
\epsfxsize=2.75in
\epsffile{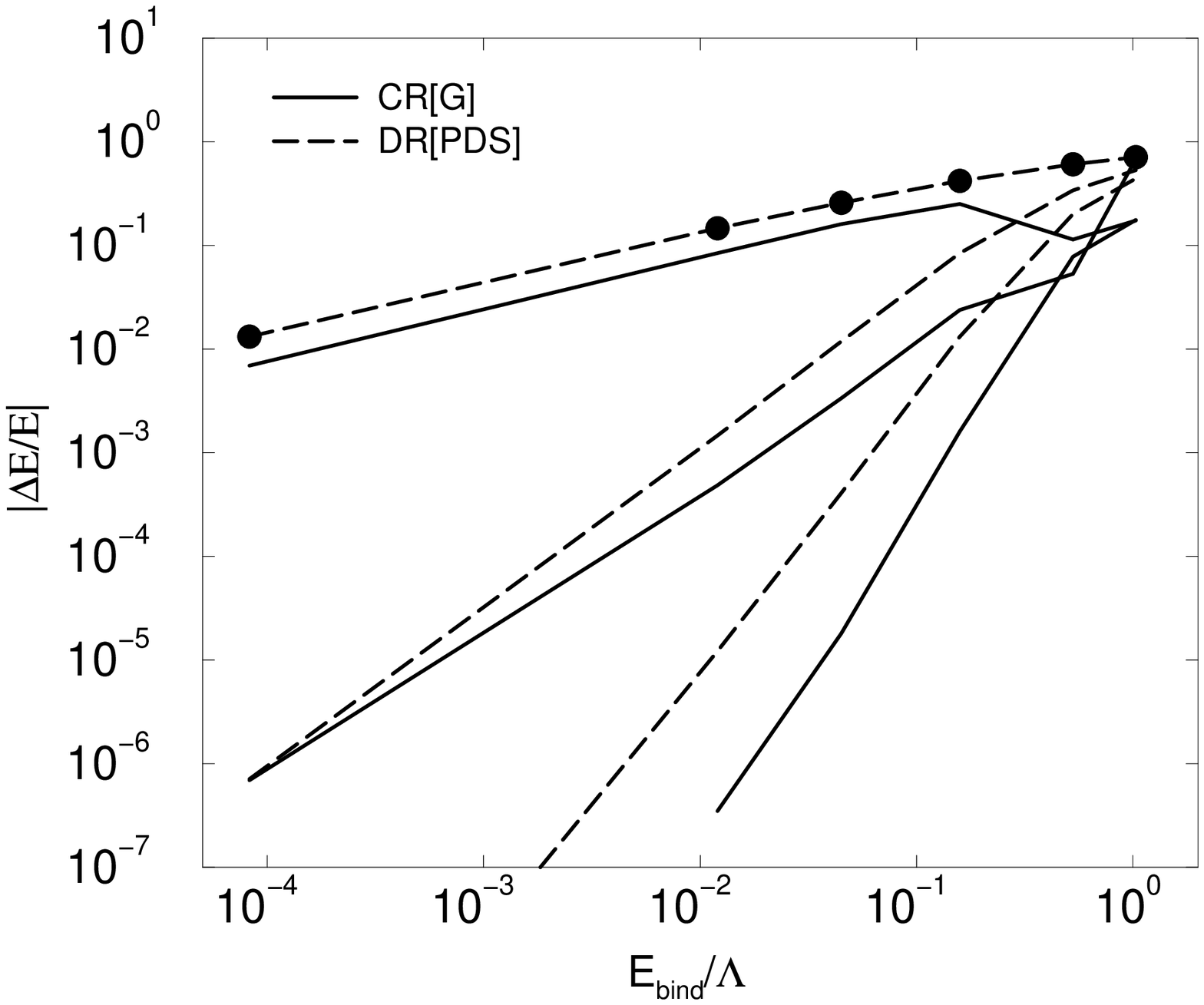}
}
\end{center}
\caption{\label{bind}The error in the binding energy for the candidate
effective field theories for representative values of $g$ from $1.01$
to $2.5$.  
The star signifies the absence of real binding energies for 
\protect\drms{} with two
constants at the values of $g>1.25$ considered.}
\end{figure}

The relative error in the binding energy is plotted in Fig.~\ref{bind}.
Both CR[G] and DR[PDS] show the clear power counting behavior and
proper radius of convergence expected from a true effective field
theory.  This gives a graphical verification that the errors in the
binding energy really do follow power counting rules.
I have checked that the same behavior is seen when plotted as a
function of the average momentum $\sqrt{\langle p^2\rangle}/\Lambda$.
The dibaryon result also follows the expected error scaling.
In contrast, the deficiencies of \drms{} regularization
seen for the phase shifts
are manifested here as binding energies that do not follow the EFT
error scaling, improving to a lesser degree than expected.  
In addition, for values of $g > 1.25$ with two
constants, the \drms{} S-matrix shows no bound state with
a real energy.

Therefore, these results show that most 
regularization procedures demonstrate
the characteristics of a systematic predictive effective
field theory.  The fit of more and more constants in the effective
potential improves the predictive power order-by-order in the momentum
expansion.  The radius of convergence of the EFT is independent of the
scattering 
length and is given by the scale where new physics enters.  

\begin{figure}
\begin{center}
\leavevmode
\hbox{
\hspace{-1.25cm}
\epsfxsize=2.75in
\epsffile{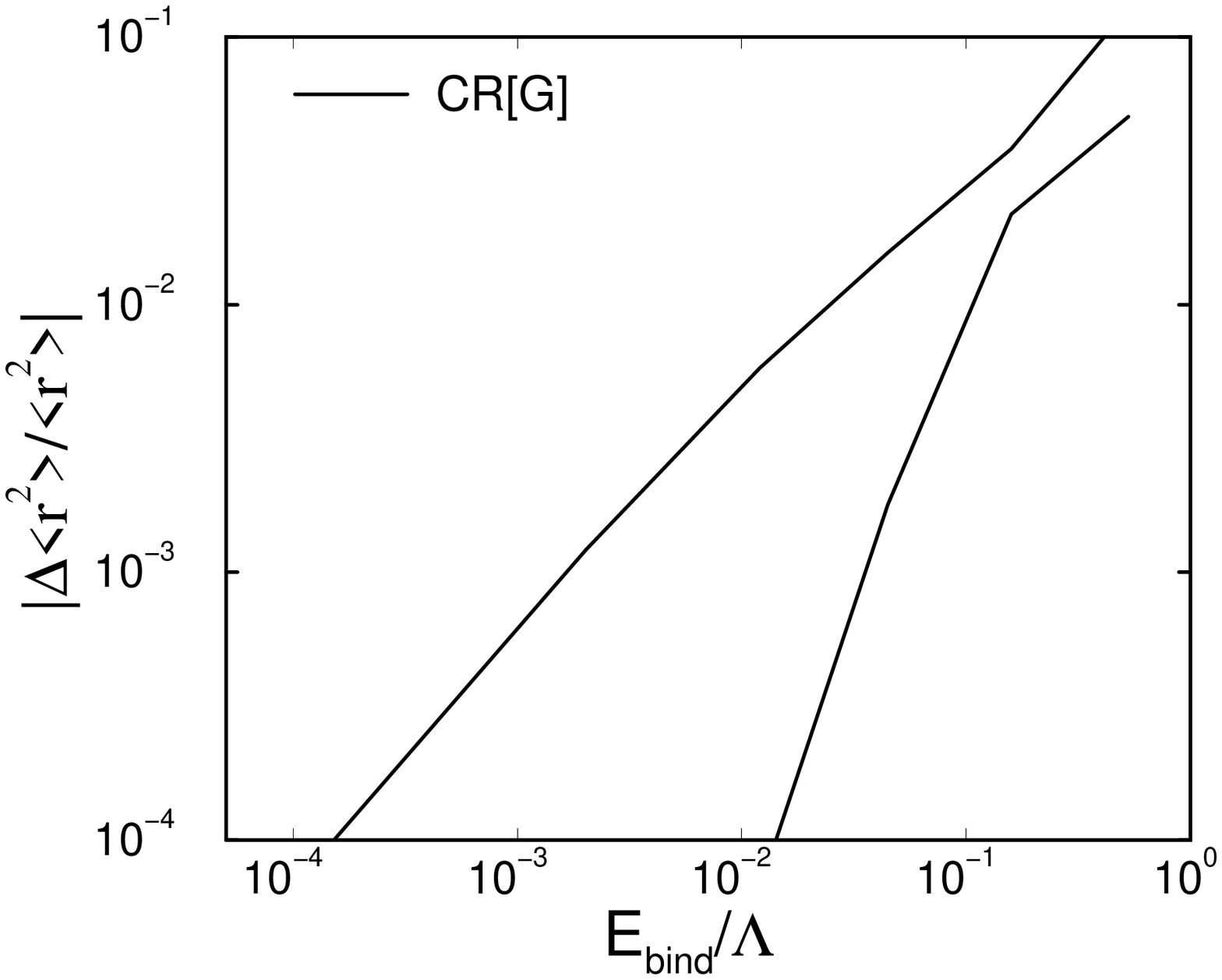}
\hspace{-.5cm}
\epsfxsize=2.75in
\epsffile{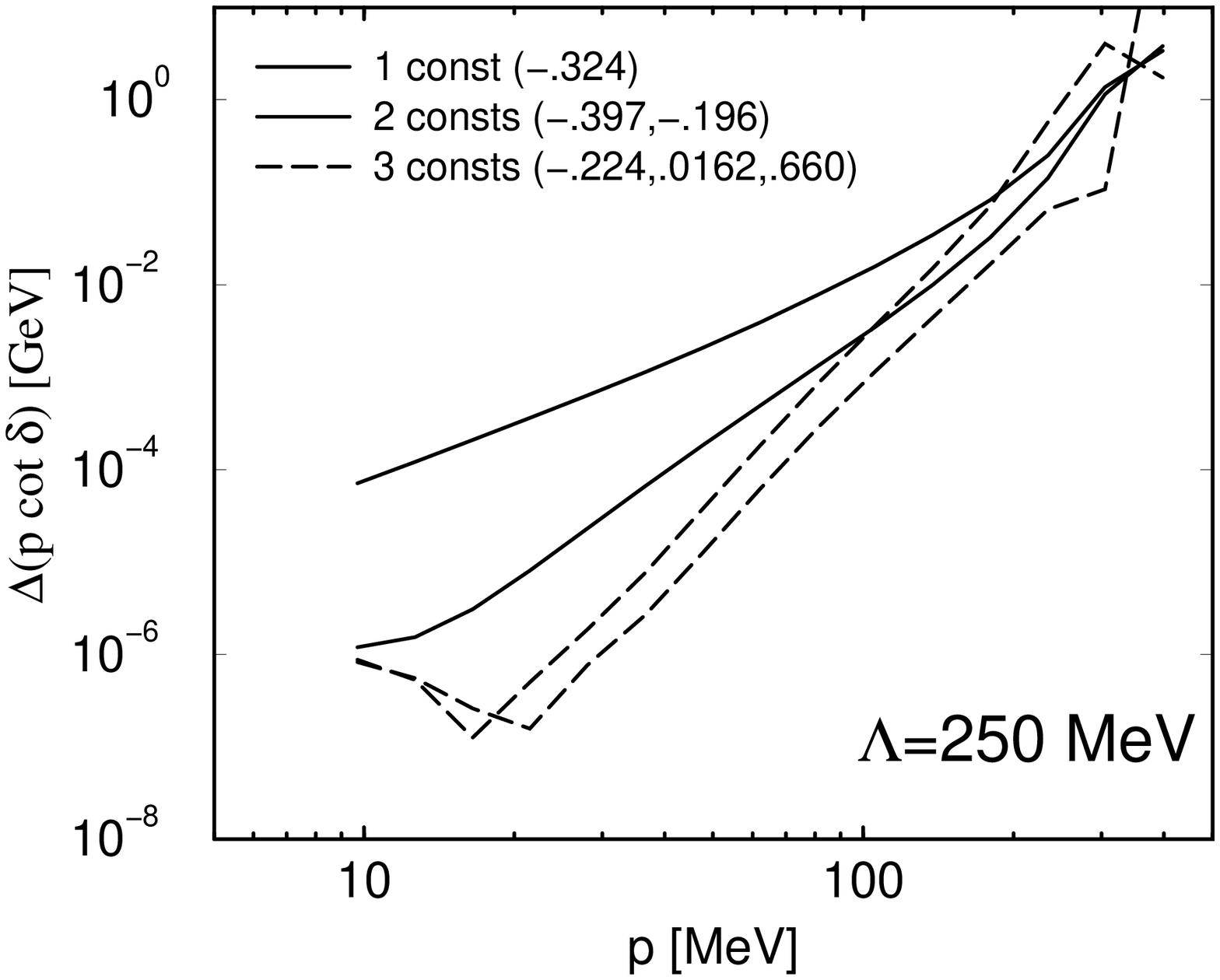}
}
\end{center}
\caption{\label{prelim}The left plot shows the error from the bound
state moment $\langle r^2 \rangle$ in the delta-shell potential.  The
right plot shows the result for one, two, and three constants in CR[G]
when fit to the $^1S_0$ $NN$ scattering data from
Nijmegen.~\protect\cite{Nij}}
\end{figure}

This test can be taken further by looking at moments of the bound state
wavefunction, such as $\langle r^2 \rangle$.  Preliminary results are
shown in the left plot of Fig.~\ref{prelim} for one and two
constants.  Going to third order is not productive here because the
effective operator $r^2$ changes~\cite{Lepage} due to the operator product
expansion 
at this order in $1/\Lambda^2$. A systematic improvement of the
expectation value is nevertheless seen for one and two constants.  The
radius of convergence does not look optimal, but this is merely an
illusion since CR[G] with one constant does slightly better than
anticipated for large $E_{\rm bind}/\Lambda$, as seen from
Fig.~\ref{bind}.

Finally, it would be ideal to apply the present analysis to real data
for $NN$ scattering.  Using the data from the Nijmegen phase shift
analysis,~\cite{Nij} I have taken the theoretical error in the EFT
analysis with CR[G] and the experimental uncertainties into account in
constructing the right plot of Fig.~\ref{prelim}.  Since two-pion
physics was not included in the analysis, the 
cutoff was taken near the $2m_\pi$ threshold ($\Lambda=250$ MeV).  The
one and two constant results show the systematic improvement and nice
radius of convergence expected from above.  However, the uncertainty
in the long distance physics blurs the three constant result given
by the range between the two dotted lines.  This must be looked at in
more detail to achieve a better precision fit.

\section{Conclusion}

I have presented results~\cite{OurPaper} showing how the error
analysis suggested by Lepage~\cite{Lepage} 
can be applied to compare cutoff regularization, two forms of 
dimensional regularization, and the dibaryon approach in the context
of nonperturbative, nonrelativistic effective field theories.
This analysis focuses on a key signature of EFT behavior: the
systematic scaling of errors with momentum or energy.

New numerical procedures~\cite{OurPaper} allow for an analysis to third
order and beyond 
in the EFT expansion, which is necessary to obtain a clear
graphical determination of the radius of convergence for a given
observable.
Such an analysis is required for a systematic fit to data regardless
of the regularization scheme. 

It is found that all of the regularization methods except for
dimensional regularization with modified minimal subtraction
are consistent with basic features expected from a useful effective
field theory:
\begin{itemize}
  \item Each additional order in the potential leads to a systematic
  improvement in the amplitude.
  \item The radius of convergence for this improvement, when
  optimized, is dictated by
  the scale of new physics.
  \item Other observables are predicted with the same accuracy as the
  amplitude at each successive improvement.
\end{itemize}
These results are consistent with the analysis of van Kolck that, with
proper resummations, any effective field theory for short-range
interactions is equivalent to an effective range expansion to the same
order.~\cite{vanKolck}

The CR[G] and DR[PDS] regularization schemes are each suitable 
for developing effective field theories of many-nucleon systems.
In future work both schemes will be used
in extending this fitting procedure and error analysis
to nuclear matter.

\section*{Acknowledgments}
This work was supported by the 
National Science Foundation
under Grants No.\ PHY--9511923 and PHY--9258270.


\section*{References}
\begin{thebibliography}{99}

\bibitem{Weinberg}S. Weinberg, \Journal{\PLB}{251}{288}{1990};
\Journal{\NPB}{363}{3}{1991}.
%
\bibitem{BiraThesis}U.~van Kolck, Ph.D. thesis, U. Texas, Aug. 1993;
C. Ordonez, L.~Ray, 
and U.~van~Kolck, \Journal{\PRL}{72}{1982}{1994};
\Journal{\PRC}{53}{2086}{1996}. 
%
\bibitem{Maryland}S.~R. Beane, T.~D.~Cohen, and D.~R.~Phillips, 
\Journal{\NPA}{632}{445}{1998}; 
D.~R.\ Phillips and T.~D.\ Cohen, \Journal{\PLB}{390}{7}{1997};
D.~R.\ Phillips, S.~R.\ Beane, and T.~D.\ Cohen, 
\Journal{\AP}{263}{255}{1998}; 
K. A. Scaldeferri, D.~R.\ Phillips, C.-W.\ Kao, and T.~D.\ Cohen,
	\Journal{\PRC}{56}{679}{1997}.
%
\bibitem{Lepage}G.~P.\ Lepage, ``How to Renormalize the Schrodinger
Equation,'' nucl-th/9706029.
%
\bibitem{KSW1}D.~B.\ Kaplan, M.~J.\ Savage, and M.~B.\ Wise,
\Journal{\NPB}{478}{629}{1996}. 
%
\bibitem{LukeManohar} M.~Luke and A.~V.\ Manohar,
\Journal{\PRD}{55}{4129}{1997}.
%
\bibitem{Kaplan}D.~B.\ Kaplan, \Journal{\NPB}{494}{471}{1997};
   P. F. Bedaque and U. van Kolck, ``Nucleon-Deuteron Scattering from
   an Effective Field Theory,'' nucl-th/9710073;
   P. F. Bedaque, H.--W.\ Hammer, and U. van Kolck, ``Effective Theory
   for Neutron-Deuteron Scattering:  Energy Dependence,'' nucl-th/9802057.
%
\bibitem{Richardson}K. G. Richardson, M.~C.\ Birse, and J.~A.\
McGovern, ``Renormalization and Power Counting in Effective
Field Theories for Nucleon-Nucleon Scattering,'' hep-ph/9708435.
%
\bibitem{Gege}J.\ Gegelia, ``Chiral Perturbation Theory Approach to
NN Scattering Problem,'' nucl-th/9802038, 1998.
%
\bibitem{OurPaper}
J. V. Steele and R. J. Furnstahl, ``Regularization Methods for
Nucleon-Nucleon Effective Field Theory,'' nucl-th/9802069.
%
\bibitem{Reid}R.~V.~Reid, \Journal{\AP}{50}{411}{1968}.
%
\bibitem{KSW2}D.~B.\ Kaplan, M.~J.\ Savage, and M.~B.\ Wise,
``A New Expansion for Nucleon-Nucleon Interactions,'' nucl-th/9801034;
``Two-nucleon Systems from Effective Field Theory,'' 
nucl-th/9802075. 
%
\bibitem{vanKolck}U. van Kolck, ``Nucleon-Nucleon
Interaction and Isospin Violation,'' hep-ph/9711222.
%
\bibitem{Gottfried}K. Gottfried, {\it Quantum Mechanics\/}, 
     (Benjamin, Reading, MA, 1979).
%
\bibitem{berger}J. R. Bergervoet {\it et al.}, in {\it Quarks and
Nuclear Structure\/}, ed.~K. Bleuler, (Springer-Verlag, New York, 1984).
%
\bibitem{ODE}These numerical routines can be found online at the Guide
to Available Math Software website http://gams.nist.gov/
%
\bibitem{MINF}M.~J.~D.\ Powell, \Journal{\CJ}{7}{155}{1964};
     W.~I.\ Zangwill, \Journal{\CJ}{10}{293}{1967}.
%
\bibitem{Nij}Partial wave data and Papers of the Nijmegen University
group can be found at http://nn-online.sci.kun.nl/.
%
\end{thebibliography}
\end{document}